\documentclass{Odyssey2026}

\usepackage{xcolor} 
\usepackage[acronym]{glossaries} 
\usepackage{cleveref} 
\usepackage{csquotes} 
\usepackage{pifont}
\usepackage{multirow} 
\usepackage{pgf}
\usepackage{pgfplots}
\usepackage{caption}
\usepackage{subcaption} 
\usepackage{rotating} 

\definecolor{yellow}{RGB}{255, 198, 0}
\definecolor{orange}{RGB}{255, 130, 0}
\definecolor{blue}{RGB}{0, 32, 91}
\definecolor{red}{RGB}{198, 53, 39}
\definecolor{magenta}{RGB}{138, 27, 97}
\definecolor{lightblue}{RGB}{0, 159, 223}
\definecolor{green}{RGB}{0, 155,119}
\definecolor{lightgreen}{RGB}{132, 189,0}
\definecolor{greenyellow}{RGB}{208, 223,0}

\definecolor{pyBlue}{rgb}{0.282352941176471,0.470588235294118,0.815686274509804}
\definecolor{pyOrange}{rgb}{0.933333333333333,0.52156862745098,0.290196078431373}
\definecolor{pyPurple}{rgb}{0.584313725490196,0.423529411764706,0.705882352941177}
\definecolor{pyGreen}{rgb}{0.415686274509804,0.8,0.392156862745098}
\definecolor{pyRed}{rgb}{0.83921568627451,0.372549019607843,0.372549019607843}
\definecolor{pyBrown}{rgb}{0.549019607843137,0.380392156862745,0.235294117647059}

\usepackage{amsmath}
\usepackage{amssymb}
\usepackage{bbm}
\usepackage{physics}
\usepackage{trfsigns}

\newcommand{\vect}[1]{\ensuremath{\boldsymbol{\mathbf{#1}}}}

\usepackage{calc}
\newcounter{z}

\colorlet{optionalColor}{black}

\newcommand{\transpose}{\textcolor{optionalColor}{\mathsf T}}

\newcommand{\yes}{\ding{51}}
\newcommand{\no}{\ding{55}}

\DeclareRobustCommand{\minwidthbox}[2]{%
\ifmmode
\expandafter\mathmakebox
\else
\expandafter\makebox
\fi
[\ifdim#1<\width\width\else#1\fi]{#2}%
}



\usepackage{etoolbox}
\renewcommand{\bfseries}{\fontseries{b}\selectfont}
\robustify\bfseries
\newrobustcmd{\B}{\bfseries}

\usepackage{tikz}
\usepackage{amssymb}
\usetikzlibrary{calc}
\usetikzlibrary{positioning}
\usetikzlibrary{chains}
\usetikzlibrary{fit}
\usetikzlibrary{arrows}
\usetikzlibrary{decorations.pathreplacing}
\usetikzlibrary{calligraphy}
\usetikzlibrary{arrows.meta}
\usetikzlibrary{backgrounds}
\usetikzlibrary{shadows}
\usetikzlibrary{tikzmark}
\usetikzlibrary{shapes.multipart}
\usetikzlibrary{matrix}

\pgfdeclarelayer{background}
\pgfdeclarelayer{foreground}
\pgfsetlayers{background,main,foreground}

\tikzset{
    every text node part/.style={align=center},
    >=stealth,
    element/.style={
        draw=black!100,
        thick,
    },
    block/.style={
        element,
        rectangle,
        minimum width=4.5em,
        minimum height=2.5em,
        inner sep=0pt,
    },
    wide block/.style={
        block,
        minimum width=7.5em,
        minimum height=2.5em,
    },
    parameter/.style={
        element,
        rectangle,
        minimum width=2.5em,
        minimum height=2.5em,
        inner sep=0pt,
        text height=1.5ex,
        text depth=.25ex
    },
    random/.style={
        element,
        circle,
        minimum width=2.5em,
        minimum height=2.5em,
        inner sep=0pt,
        draw=black!100,
        text height=1.5ex,
        text depth=.25ex
    },
    observation/.style={
        random,
        double
    },
    branch/.style={
        circle,
        fill=black,
        draw=black,
        minimum size=0.2em,
        inner sep=0pt
    },
    apply/.style={
        circle,
        thick,
        draw=black!100,
        minimum size=1em,
        inner sep=0pt,
        label=center:{$\times$}
    },
    node distance=2em,
    arrow/.style={->,shorten >=0.1em},
    reverse arrow/.style={<-,shorten <=0.1em},
}

\tikzset{pics/.cd,
pic switch/.style args={#1 times #2}{code={
\tikzset{x=#1/2,y=#2/2}
\coordinate (-north west) at (-1,1);
\coordinate (-north east) at (1,1);
\coordinate (-south west) at (-1,-1);
\coordinate (-south east) at (1,-1);
\coordinate (-north) at (0,1);
\coordinate (-east) at (1,0);
\coordinate (-south) at (0,-1);
\coordinate (-west) at (-1,0);
\coordinate (-in) at (1,0);
\coordinate (-closed) at (-1,1);
\coordinate (-opened) at (-1,-1);

\draw [line cap=rect] (-1,1) -- (-1,0.5);
\draw [line cap=rect] (-1,-1) -- (-1,-0.5);
\draw [line cap=round] (1, 0) -- ($(1, 0)!2/3!(-1.3,0.8)$);
\draw [line cap=rect] ($(1, 0)!1/3!(-1.3,0.8)$) -- (-1.3,0.8);
}}
}

\newcommand\speechsignal[1]{\ensuremath{s^{\textrm{#1}}(t)}}

\newcommand\featureseq{\vect{X}}
\newcommand\feature{\vect{x}}

\newcommand\framescore{\ensuremath{q}}

\newcommand\uttscore{\ensuremath{y}}
\newcommand\classlabel{\ensuremath{c}}
\newcommand\contrastembedding{\ensuremath{\vect{z}}}

\newcommand{\seq}[2]{%
    \ensuremath{%
        [#1_1, \dots, #1_{#2}]
    }
}

\newcommand\encA{\ensuremath{\operatorname{Enc}}}

\newcommand\decA{\ensuremath{\operatorname{Dec}^\textrm{MOS}}}
\newcommand\decB{\ensuremath{\widetilde{\operatorname{{Dec}}}^\textrm{MOS}}}

\newcommand\decsclA{\ensuremath{\widetilde{\operatorname{Dec}}^{\textrm{scl}}}}

\newcommand\proj[2]{%
    \ensuremath{\operatorname{Proj}^{\textrm{#1}}\left(#2\right)}
}

\newcommand\relu[1]{\ensuremath{\max\left(#1, 0\right)}}
\newcommand{\pmixup}{\ensuremath{p_{\textrm{mixup}}}}

\newcommand{\formatEER}[1]{$#1$}
\newcommand{\formatEERB}[2]{\formatEER{#1}&\formatEER{#2}}

\newcommand\formatEERD[3]{\formatEER{#1}&\formatEER{#2}&\formatEER{#3}}
\newcommand\formatAUC[1]{$#1$}

\newcommand\formatAUCD[3]{\formatAUC{#1}&\formatAUC{#2}&\formatAUC{#3}}
\newcommand{\IAUC}[2]{$\textrm{I}_{#1,#2}$-AUC}

\newacronym{mos}{MOS}{mean opinion score}
\newacronym{sqa}{SQA}{%
    speech quality assessment
}
\newacronym{ssqa}{SSQA}{%
    subjective \acrlong{sqa}
}
\newacronym{lssqa}{LSSQA}{%
    local \acrlong{ssqa}
}
\newacronym{ssl}{SSL}{%
    self-supervised learning
}
\newacronym{sed}{SED}{%
    sound event detection
}

\newacronym{eer}{EER}{equal error rate}
\newacronym{dcf}{DCF}{detection cost function}
\newacronym{mindcf}{minDCF}{minimum detection cost function}
\newacronym{auc}{AUC}{area under curve}
\newacronym{iauc}{I-AUC}{intersection-based area under curve}
\newacronym{ari}{ARI}{adjusted rand index}

\makeatletter
\def\input@path{{/}{figures/}}
\makeatother

\hypersetup{
  breaklinks=true,
}

\interspeechcameraready 

\title{%
    Speech Quality Embeddings for Improved Detection and Classification\\ of Degradations in Speech Signals
}

\author[affiliation={1}]{Michael}{Kuhlmann}
\author[affiliation={1}]{Tobias}{Cord-Landwehr}
\author[affiliation={1}]{Reinhold}{Haeb-Umbach}

\affiliation{}{Paderborn University}{Germany}
\email{\{kuhlmann,haeb\}@nt.uni-paderborn.de}

\usepackage{comment}
\usepackage{soul}
\begin{document}
\ninept
\abovedisplayskip5pt plus 3.0pt minus 4.0pt
\belowdisplayskip5pt plus 3.0pt minus 4.0pt

\setlength{\floatsep}{1ex}
\setlength{\textfloatsep}{1em}

\setlength{\abovedisplayskip}{3pt}
\setlength{\belowdisplayskip}{3pt}
\setlength{\textfloatsep}{10pt plus 0.0pt minus 2.0pt}

\setlength{\floatsep}{10pt plus 0.0pt minus 3.0pt}
\setlength{\intextsep}{10pt plus 0.0pt minus 3.0pt}

\maketitle

\begin{abstract}
    Automatic subjective speech quality assessment (SSQA) traditionally estimates speech quality on an utterance or system level. While this resolution was adequate for older transmission or synthesis systems that produced speech signals of mediocre quality, modern systems generate high-quality speech with degradations that may occur only locally. With suitable model architectures and regularization losses, SSQA models trained with utterance-level targets can also yield useful local predictions of speech quality. In this work, we extend such models to produce frame-level embeddings that cluster by degradation type. Specifically, we employ a partial mix-up strategy on a parallel corpus of clean and degraded utterances and apply a contrastive loss to distinguish between degradation types. Through experiments on both in- and out-of-domain data, we demonstrate that our approach improves degradation detection and enables the identification of degradation types by analyzing embedding clusters.
\end{abstract}

\section{Introduction}

Automatic \acrfull{ssqa} assigns a quality score to speech signals that reflects subjective quality perception (e.g., poor or excellent)~\cite{huang2025sheet}.
Usually, these approaches work \emph{non-intrusively}, i.e., without knowledge of a clean matching reference signal.
Furthermore, they are trained with \gls{mos} supervision - gathered from listening experiments - and provide a single score for the complete utterance (\emph{global} score).
However, the expressivity of a global score is limited and, for example, cannot inform the user about temporally constrained artifacts in the speech~\cite{kuhlmann2025towards,kuhlmann2026speech}.
Supplementing global evaluation with frame-level (\emph{local}) evaluation can help with artifact localization~\cite{kuhlmann2026speech}, increase interpretability~\cite{fu2018qualitynet,kuhlmann2025towards,kuhlmann2026speech}, and may be used as a learning signal to reduce such artifacts in future model iterations~\cite{fu2019learning,fu2019metric}.
Beyond localizing artifacts, automatic identification of artifact types can further increase interpretability and retrieval of specific artifacts~\cite{cumlin2025impairments}.


Only a few prior works have looked into extending \gls{sqa}\footnote{Following~\cite{huang2025sheet}, we use the term \acrlong{sqa} to denote the broader class of models that are not only trained with human preference data but with signal-based intrusively-derived labels like PESQ.} models to include these capabilities.
Quality-Net~\cite{fu2018qualitynet} trained an \acrshort{sqa} model with frame-level regularization to predict plausible frame-level quality scores.
Recently, frame-level regularization was applied to stronger \acrshort{ssqa} models based on \gls{ssl}~\cite{kuhlmann2025towards,kuhlmann2026speech}.
To detect local degradations, they tuned and applied a threshold on the predicted frame-level \acrshort{mos} (\emph{\acrshort{mos}-based degradation detection}).
Because no training-scale data with frame-level quality scores exist, all these approaches rely on utterance-level targets from which frame-level quality scores should be learned.

Learning to detect regions of signal degradation from a global, i.e., \emph{weak}, target shows large resemblances to the research field of \gls{sed}~\cite{turpault2019sed,hao2022dcase,chen2023dcase}.
Recently, frame-level and time-aligned audio-language models~\cite{wu2025flam,primus2025tacos} were developed that replace the static set of known classes commonly used in many \acrshort{sed} approaches with textual descriptions.
To do so, these approaches share two key factors: training with frame-level (\emph{strong}) (pseudo-)targets and using a contrastive loss to align frame-level embeddings with text annotations.

On the frontier of identifying degradations, Cumlin et al.~\cite{cumlin2025impairments} found that the latent space of DNSMOS-like~\cite{reddy2021dnsmos} models clusters by degradation type.
They used a kNN classifier to identify known impairments and could further improve the results by training on impairment-augmented data with PESQ~\cite{rix2001perceptual} or ViSQOL~\cite{chinen2020visqol} labels.
However, their observation holds for utterance-averaged embeddings, in which impairments dominate the speech signal, and it is unclear whether it generalizes to local degradations.
NOMAD~\cite{ragano2024nomad} used a triplet loss between degraded signals to learn perceptual embeddings.
As positive pairs were constructed using close NSIM~\cite{hines2012speech}, the embeddings capture degradation intensity rather than type.

In this work, we show that the non-intrusive \emph{joint detection and identification} of local degradations in a speech signal can be learned by borrowing concepts from frame-wise audio language models, namely using data augmentation to generate strong \emph{pseudo-targets}~\cite{wu2025flam} and using a contrastive loss to align similar concepts in the embedding space~\cite{wu2025flam,primus2025tacos}.
Our contributions are twofold.
First, using a parallel data corpus of clean reference and degraded speech signals with utterance-level \acrshort{mos}-scores, we propose a \emph{partial mix-up} strategy to generate training data with local degradations.
Using frame-level pseudo-targets from a pretrained  \acrshort{ssqa} model~\cite{kuhlmann2026speech}, we train a new frame-level \acrshort{ssqa} model on these mix-up data.
Compared to the baseline~\cite{kuhlmann2026speech}, \acrshort{mos}-based degradation detection can be significantly improved when adding this frame-level supervision during training.
Second, to improve the similarity of embeddings derived from similarly degraded frames, we build upon the partial mix-up augmentation strategy and add a frame-level supervised contrastive loss~\cite{khosla2020supervised}, exploiting knowledge of the degradation classes used during mix-up.
We find that switching to an embedding-based degradation detection method, using a clean reference embedding as the enrollment, yields the best detection performance.
Finally, an analysis of the latent embedding space reveals that different degradation types form distinct clusters, but that cluster purity degrades under multiple simultaneous degradations or when switching to out-of-domain degradations.

\section{Local speech quality assessment}
\Acrfull{lssqa}\footnote{In the following, \acrshort{sqa}/\acrshort{ssqa} always refer to the \emph{automatic} quality assessment of speech.} aims to estimate quality at a finer resolution than the utterance level.
Similar to utterance-level assessment, encoder-decoder models~\cite{huang2025sheet} can be used to infer frame-level quality scores\footnote{In the following, we will refer to these as \emph{frame-level scores}.}~\cite{kuhlmann2025towards}.

Given a dataset $\mathcal{D}=\{(s_1(t),y_1),(s_2(t),y_2),\dots\}$ of speech signals $s_i(t)$ with utterance-level \acrshort{mos} annotation $y_i$, the frame-level scores are inferred from an audio signal as follows. 
First, $s(t)$ is processed by an encoder, which extracts frame-level embeddings
${\featureseq = \encA(s(t))=[\feature_1, \dots, \feature_l, \dots, \feature_L]}$ with frame index $l$.
Then, a decoder network is applied to the frame-wise audio embeddings, estimating the \acrshort{mos} of each frame as ${\vect{\hat{\framescore}} = \decA(\featureseq)=[\hat{\framescore}_1, \dots, \hat{\framescore}_l, \dots, \hat{\framescore}_L]}$.
Since frame-level targets are usually unavailable during training time, approaches rely only on utterance-level targets~\cite{cooper2021generalization,mittag2021nisqa,saeki2022utmos}, make specific architectural choices~\cite{kuhlmann2025towards}, and add regularization losses during training to infer frame-level scores~\cite{fu2018qualitynet,kuhlmann2026speech}.
In~\cite{kuhlmann2025towards}, a conventional \acrshort{ssqa} loss~\cite{leng2021mbnet,saeki2022utmos} with a pool-last strategy was used
\begin{align}
    &\mathcal{L}_{\textrm{\acrshort{ssqa}}} =\\
    &\quad \frac1B\sum_{b=1}^B \relu{\left\lvert\uttscore_b-\frac1L\sum_{l=1}^L\hat{\framescore}_{b,l}\right\rvert-\delta}
    +\mathcal{L}^{\textrm{\acrshort{mos}-con}}, \nonumber
\end{align}
where $y_b$ and $\hat{\framescore}_{b,l}$ denote the target \acrshort{mos} and estimated $l$-th frame \acrshort{mos} of the $b$-th example in a mini-batch of size $B$, $\delta$ is a clipping parameter~\cite{leng2021mbnet} and $\mathcal{L}^{\textrm{\acrshort{mos}-con}}$ is a contrastive loss on the utterance-level scores~\cite{saeki2022utmos}.
In~\cite{kuhlmann2026speech}, an additional consistency loss between data slices was added to reduce the influence of far-away context frames on the embeddings and scores
\begin{align}
    \label{eq:loss-lssqa}
    &\mathcal{L}_\textrm{\acrshort{lssqa}} = \\
    &\quad \mathcal{L}_{\textrm{\acrshort{ssqa}}} + \frac{1}{\Delta{}L}\left(\sum_{l=l_0}^{l_0+\Delta{}L}\lVert\feature_l-\feature^{\textrm{slice}}_l\rVert_2^2
    + \lvert\hat{\framescore}_l-\hat{\framescore}^{\textrm{slice}}_l\rvert\right), \nonumber
\end{align}
where $\feature^{\textrm{slice}}$ and $\hat{\framescore}^{\textrm{slice}}$ denote encoder embeddings and frame-level scores that are extracted from an input signal slice $\speechsignal{slice}$ of duration $\Delta{}L$ frames (see \cite{kuhlmann2026speech} for details).

\section{Improved frame-level scores}
\label{sec:improving}
Local degradation detection is challenging because only utterance-level, i.e., weak, quality targets are available at scale.
As motivated at the beginning, we expect that training with frame-level targets will improve detection performance.
To this end, we propose a straightforward data augmentation strategy to produce frame-level labels for training an \acrshort{lssqa} model by defining pseudo-targets. 
These pseudo-targets are defined using a corpus of parallel data comprising clean reference and degraded speech utterances, $\mathcal{D}_{\textrm{ref}}$ and $\mathcal{D}_{\textrm{deg}}$, respectively\footnote{Such corpora can be obtained by simulating signal distortions~\cite{cumlin2025impairments,mittag2021nisqa}, voice conversion, or forced-aligned text-to-speech models.}.
We employ a pretrained \acrshort{lssqa} model~\cite{kuhlmann2026speech} to obtain parallel sets of frame-level scores, $\mathcal{Q}_\textrm{ref}$ and $\mathcal{Q}_{\textrm{deg}}$, from $\mathcal{D}_{\textrm{ref}}$ and $\mathcal{D}_{\textrm{deg}}$, respectively.
During training, we sample for each audio with probability $\pmixup$ binary mix-up masks $m(t)$ and ${\vect{m}_\framescore=\seq{m}{L}}$, with $m(t),m_l\in\{0, 1\}$, to generate augmented training data with frame-level pseudo targets
\begin{align}
    \speechsignal{\textrm{pseudo}} &= m(t)\speechsignal{\textrm{deg}} + (1-m(t))\speechsignal{\textrm{ref}}, \\
    \vect{{\framescore}}^{\textrm{pseudo}} &= \vect{m}_\framescore\odot\vect{\hat{\framescore}}^{\textrm{deg}} + (1-\vect{m}_\framescore)\odot\vect{\hat{\framescore}}^{\textrm{ref}},
\end{align}
where $m(t)$ and $\vect{m}_\framescore$ are defined in sample and frame resolution, respectively, and otherwise mask the same time points, and $\odot$ denotes element-wise multiplication.
Importantly, for this to work, we require that the degradations persist throughout the complete utterance taken from $\mathcal{D}_{\textrm{deg}}$ with a similar strength.
Then, it suffices that the pretrained model can predict frame-level scores and has a high utterance-level correlation with the target \acrshort{mos}; it need not be fine-tuned for detection.
We then train a new \acrshort{lssqa} model with additional frame-level supervision based on the pseudolabels $\vect{{\framescore}}^{\textrm{pseudo}}$
\begin{align}
    \label{eq:loss-lssqa-sup}
    \mathcal{L}^{\textrm{sup}}_{\textrm{\acrshort{lssqa}}} =
    \quad\mathcal{L}_{\textrm{\acrshort{lssqa}}} + \frac1{BL}\sum_{b=1}^B\sum_{l=1}^L\lvert\hat{\framescore}_{b,l}-\framescore^{\textrm{pseudo}}_{b,l}\rvert,
\end{align}
where $\vect{\hat{\framescore}}$ is now inferred as $\decA\left(\encA\left(\vect{s}^{\textrm{pseudo}}\right)\right)$ and we set $y_b=\frac1L\sum_{l=1}^L\framescore^{\textrm{pseudo}}_{b,l}$ in $\mathcal{L}_{\textrm{\acrshort{ssqa}}}$ when partial mix-up was applied.

\section{Speech quality embeddings}
For downstream applications, not only the location of a degradation in a signal, but also its \emph{type} is relevant.
This can be beneficial in system analysis, e.g., identifying frequent degradation types or retrieving specific degradations from large databases.
Cumlin et al.~\cite{cumlin2025impairments} studied the latent space of DNSMOS-like~\cite{reddy2021dnsmos} models and found that the utterance-level embeddings form clusters by degradation type.
However, unlike DNSMOS, we perform time pooling at the decoder output, and we find that neither our encoder nor decoder embeddings cluster well by degradation type.
Moreover, Deng~\cite{deng2025investigating} recently showed that audio embeddings do not always form clear clusters under different degradations and degradation strengths.

\subsection{Contrastive learning for quality embeddings}

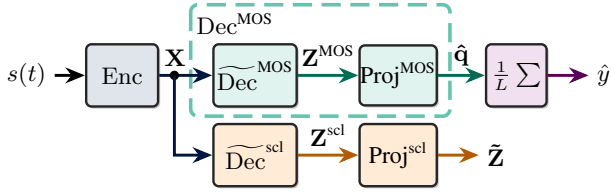
\begin{figure}[t]
    \tikzset{
    circ/.style = {
		draw,
		thick,
		circle,
		inner sep = 0,
		minimum width = #1,
		fill = black,
	},
}

\begin{tikzpicture}[
    node distance=0.62cm and 0.54cm,
    font=\small,
    fancybox/.style={
        block,
        rounded corners=2pt,
        line width=0.9pt,
        draw=black!85,
        drop shadow={shadow xshift=0.8pt,shadow yshift=-0.8pt,opacity=0.25},
        drop shadow={shadow xshift=0.8pt,shadow yshift=-0.8pt,opacity=0.25}
    },
    encblock/.style={fancybox, fill=blue!12},
    mosblock/.style={fancybox, fill=green!12},
    embblock/.style={fancybox, fill=orange!16},
    aggblock/.style={fancybox, fill=violet!12},
    arrbase/.style={-{Stealth[length=2.6mm,width=2.1mm]}, line width=1.1pt},
    encarr/.style={arrbase, draw=blue!70!black},
    mosarr/.style={arrbase, draw=green!70!black},
    embarr/.style={arrbase, draw=orange!70!black},
    aggarr/.style={arrbase, draw=violet!70!black}
]
    \node[] (input) {$s(t)$};
    \node[encblock, right=0.4cm of input, minimum width=0.95cm, minimum height=0.82cm] (enc) {$\encA$};
    \node[mosblock, right=0.7cm of enc, minimum width=1.1cm, minimum height=0.82cm] (decmos) {$\decB$};
    \node[mosblock, right=0.8cm of decmos, minimum width=1.02cm] (lin) {$\textrm{Proj}^{\textrm{MOS}}$};
    \node[aggblock, right=0.62cm of lin, minimum width=0.8cm, minimum height=0.82cm] (sum) {$\frac1L\sum$};
    \node[right=of sum] (utt) {$\hat{\uttscore}$};

    \draw[draw=green!60,rounded corners=0.3em,line width=0.15em,dash pattern=on 0.8em off 0.4em] ($(decmos.north west)+(-0.9em,1.6em)$) rectangle ($(lin.south east)+(0.5em,-0.4em)$);
    \node[] at ($(decmos.north west)+(0.3,0.8em)$) {\decA};

    \node[embblock, below=0.2cm of decmos, minimum width=1.1cm, minimum height=0.82cm] (decscl) {$\decsclA$};
    \node[embblock, right=0.8cm of decscl, minimum width=1.02cm, minimum height=0.82cm] (proj) {$\textrm{Proj}^{\textrm{scl}}$};
    \node[right=of proj] (zproj) {$\vect{\tilde{Z}}$};

    \draw[arrbase] (input) -- (enc);
    \draw[encarr] (enc) -- node[above,pos=0.3] {$\featureseq$} (decmos);
    \draw[mosarr] (decmos) -- node[above,midway] {$\vect{Z}^{\textrm{MOS}}$} (lin);
    \draw[mosarr] (lin) -- node[above,midway] {$\vect{\hat{\framescore}}$} (sum);
    \draw[aggarr] (sum) -- (utt);

    \node[circ, minimum width=0.3em] at ($(enc.east)+(0.2, 0)$) {};
    \draw[encarr] ($(enc.east)+(0.2, 0)$) |- (decscl.west);
    \draw[embarr] (decscl) -- node[above] {$\vect{Z}^{\textrm{scl}}$} (proj);
    \draw[embarr] (proj) -- (zproj);
\end{tikzpicture}
    \vspace{-1.5em}
    \caption{%
        Block schema of the full proposed model.
        The encoder $\encA$ feeds two decoder heads:
        $\decA$ for frame-level scores $\vect{q}$, followed by mean pooling for utterance-level estimates $\hat{\uttscore}$, and $\decsclA$ for frame-level embeddings $\vect{Z}^{\textrm{scl}}$, followed by a projection layer for contrastive training.
        $\vect{Z}^{\textrm{MOS}}$ denotes the embeddings of the MOS decoder before projection to the frame-level scores.
    }
    \label{fig:model-overview}
\end{figure}

To learn discriminative frame-level degradation-type embeddings, we assume that the degraded dataset used for pseudo-supervised training contains annotations about the degradation type $c$: ${\mathcal{D}_{\textrm{deg}}=\{(\speechsignal{\textrm{deg}},y,c)_i\}}$~\cite{mittag2021nisqa, cumlin2025impairments}.
Using these annotations, we add a supervised contrastive loss~\cite{khosla2020supervised} to distinguish between degradation types as follows.
After applying partial mix-up, a class label $\classlabel_l\in\Omega=\{1,\dots, K+1\}$ is obtained for each frame, where $K$ is the total number of degradation classes during training, and an additional class is added to denote clean, non-degraded frames.
To decouple \acrshort{mos} estimation from representation learning, a second decoder is added to extract frame-level bottleneck embeddings ${\vect{Z}^{\textrm{scl}}=\decsclA(\featureseq)=\seq{\contrastembedding^{\textrm{scl}}}{L}}$,
${\contrastembedding^{\textrm{scl}}_l\in\mathbb{R}^{D_Z}}$.
Following~\cite{khosla2020supervised}, an additional linear projection ${\tilde{\contrastembedding}=\proj{\textrm{scl}}{\contrastembedding^{\textrm{scl}}}\in\mathbb{R}^{D_P}}$ is applied, where both $\contrastembedding^{\textrm{scl}}$ and $\tilde{\contrastembedding}$ are normalized to lie on the unit hypersphere.
\Cref{fig:model-overview} shows a schema of the full system.

Given a mini-batch of embeddings $\mathcal{\tilde{Z}}_b\in\mathbb{R}^{B\times{}L\times{}D_P}$ and corresponding frame-level class affiliations $\mathcal{C}_b\in\Omega^{B\times{}L}$, the following supervised contrastive loss is added during training:
\begin{equation}
    \label{eq:loss-supcon}
    \mathcal{L}^{\textrm{scl}} = -\frac1{BL}\sum_{b=1}^{B} \sum_{l=1}^{L} \mathcal{L}^{\textrm{scl}}_{b,l} 
\end{equation}
with
\begin{align}
    &\mathcal{L}^{\textrm{scl}}_{b,l} = \label{eq:supcon-single}\\
    &\quad\frac1{\lvert{}I_P(c_{b,l})\rvert{}}\sum_{\substack{(b^\prime,l^\prime)\in{}I_P(c_{b,l})\\(b^\prime,l^\prime)\neq(b,l)}}\log \frac{\exp\left(\tilde{\contrastembedding}_{b,l}^{\transpose}\tilde{\contrastembedding}_{b^\prime,l^\prime}/\tau\right)}{\displaystyle{\sum_{\tilde{b}}\sum_{\substack{\tilde{l}\\(\tilde{b},\tilde{l})\neq(b,l)}}\exp\left(\tilde{\contrastembedding}^{\transpose}_{b,l}\tilde{\contrastembedding}_{\tilde{b},\tilde{l}}/\tau\right)}}, \nonumber
\end{align}
where $I_P(c_{b,l})=\{(b^\prime,l^\prime)\in\{1,\dots, B\}\times\{1,\dots, L\}: c_{b,l}=c_{b^\prime,l^\prime}\}$ is the set of indices in the mini-batch that have the same class label $c_{b,l}$ and $\tau>0$ is a temperature parameter.
Essentially, this loss computes the cross-entropy between a softmax distribution of $B\cdot{}L$ elements over $B\cdot{}L$ classes and the multi-hot target distribution matrix $\vect{P}$, where $p_{bl,b^\prime{}l^\prime}=1/\lvert{}I_P(c_{b,l})\rvert$ if $c_{b,l}=c_{b^\prime,l^\prime}$ and $0$ else~\cite{tian2023stablerep}.
Different from~\cite {khosla2020supervised}, the contrastive loss is applied to frame-level embeddings here, meaning that embeddings $\tilde{\contrastembedding}_{b,l}$ and $\tilde{\contrastembedding}_{b,l+\lambda}$ for some small value $\lambda$ exhibit strong correlation.
We consider the embeddings $\{(\tilde{\contrastembedding}_{b,l_1},\tilde{\contrastembedding}_{b,l_2}):|l_1-l_2|<\lambda\}$ as a special case of self-contrast and exclude these pairs from the summations in \cref{eq:supcon-single}, where $\lambda$ is a hyperparameter.
Incorporating the supervised contrastive loss into the training of our \acrshort{lssqa} model, the total loss becomes
\begin{equation}
    \mathcal{L}_{\textrm{total}} = \mathcal{L}^{\textrm{sup}}_{\textrm{\acrshort{lssqa}}} + \tau\mathcal{L}^{\textrm{scl}}, \label{eq:total-loss}
\end{equation}
where we follow~\cite{khosla2020supervised} and scale $\mathcal{L}^{\textrm{scl}}$ with the temperature.

\subsection{Handling non-degraded frames}
With the partial mix-up strategy, many clean, i.e., non-degraded frames, will be involved in the computation of $\mathcal{L}^{\textrm{scl}}$.
There are two possible ways to handle these frames:
i) Treat non-degraded frames as a distinct class or ii) ignore their contribution to $\mathcal{L}^\textrm{scl}$.
In the latter case, we modify \cref{eq:loss-supcon} to exclude specific classes
\begin{align}
    \mathcal{L}^{\textrm{scl}}
    = -\frac1{BL}\sum_{b=1}^B\sum_{l=1}^L \mathbbm{1}\left(c_{b,l}\notin\overline\Omega\right)\mathcal{L}_{b,l}^\textrm{scl},
\end{align}
where $\overline\Omega$ is the set of class indices that should be excluded.
Note that the summations in the denominator in~\cref{eq:supcon-single} still run over the excluded classes.
Especially in a sparse scenario, where the number of degraded frames is significantly lower than the number of clean frames, the latter case may help learn more distinct degradation clusters.

\section{Local degradation detection: MOS-based versus embedding-based}
\label{sec:localization}

Given estimated frame-level scores $\vect{\hat{\framescore}}$, \emph{\acrshort{mos}-based} detection of local degradations infers, for each frame, whether it suffers from a quality degradation by comparing its frame-level score to a score threshold $\framescore_{\textrm{deg}}$ that is tuned on a validation set.

Assuming that the embeddings form clusters by degradation type and that non-degraded frames cluster together, we can perform degradation detection with the encoder ($\featureseq$) or decoder ($\vect{Z}^{\textrm{MOS}}$, $\vect{Z}^{\textrm{scl}}$, or $\tilde{\vect{Z}}$) \emph{embeddings} as an alternative to \acrshort{mos}-based detection.
Given a clean reference embedding for enrollment, we compute the cosine similarity between the enrollment embedding and all input embeddings, similar to NOMAD~\cite{ragano2024nomad}, and compare these similarities against a threshold to identify degraded frames.
Compared to \acrshort{mos}-based detection, the detection performance depends on the choice of the enrollment and provides an easy way to tune the model to the inference data by selecting a matching enrollment.
Note that a similar strategy can be used to retrieve frames exhibiting a specific degradation when a corresponding enrollment is available, analogous to a speaker verification~\cite{snyder2018x,chung2020defence}.

\section{Experiments}
\label{sec:experiments}
To demonstrate that our contributions improve the detection of local degradations and the identification of degradation types, we conduct three evaluations.
First, the proposed \acrshort{lssqa} extensions are evaluated with respect to their \acrshort{mos}- and embedding-based degradation detection.
Then, the speech quality embeddings are assessed for their specificity in distinguishing among degradation types.
Finally, the models' performance for joint degradation detection and degradation-type clustering is investigated.




\subsection{Training \& evaluation setup}
We train all models on NISQA~\cite{mittag2021nisqa} and BVCC~\cite{cooper2021how}.
For NISQA, we apply the partial mix-up strategy as explained in~\Cref{sec:improving}.
When training with $\mathcal{L}^{\textrm{sup}}_{\textrm{\acrshort{lssqa}}}$, we sample a mask $\vect{m}$ by sampling between 1 and 3 mask segments, each of duration between $\SI{200}{ms}$ and $\SI{1}{s}$.
When training with $\mathcal{L}_{\textrm{total}}$, we impose additional constraints during mask sampling so that the mix-up regions contain perceptible degradations almost surely:
We exclude any utterances from mix-up
where the utterance-level score of the reference $y^{\textrm{ref}}<3.5$ or where the score of the degradation $y^{\textrm{deg}}>4$, we only sample from speech regions\footnote{We use rVAD~\cite{tan2020rvad} to detect speech regions.}, and we only sample from regions where ${\vect{\framescore}^{\textrm{ref}}-\vect{\framescore}^{\textrm{deg}}>0.5}$.
To construct the class labels, we read the metadata from the \texttt{NISQA\_TRAIN\_SIM} and \texttt{NISQA\_VAL\_SIM} subsets.
In NISQA, often multiple degradations were applied to a single audio file.
We treat the combination of multiple simultaneous degradations as a distinct class, as in~\cite {cumlin2025impairments}.
This yields $K=899$ distinct classes for the training split and $K=371$ for the validation split, each split using the same $19$ unique single degradations.
All data preparation steps are open-sourced\footnote{%
    \ifinterspeechfinal
        \url{https://github.com/fgnt/local_sqa}
    \else
        Hidden during review.
    \fi
}.

\subsubsection{Validation and test data}
As there exists no established benchmark dataset for local degradation detection, we use two synthetic datasets to report our results, each with a validation and test split.

\noindent\textbf{\texttt{NISQA\_VAL\_SIM}-partial-mixup} and \textbf{\texttt{NISQA\_TEST\_SIM}-partial-mixup}. These are in-domain validation and test datasets, as they include a similar set of degradations to the training data, and we use the same partial mix-up strategy as in the training to generate them, where \texttt{NISQA\_TEST\_SIM} is the combination of \texttt{NISQA\_TEST\_FOR} and \texttt{NISQA\_TEST\_P501}.
For both splits, we use slightly longer degradation segments (between $\SI{400}{ms}$ and $\SI{2}{s}$) than seen during training.
The test split contains $9$ unique single degradations and a total of $K=36$ degradation classes.

\noindent\textbf{LibriAugmented-partial-mixup}. To test generalization to out-of-domain degradations, we apply partial mix-up to the \textit{dev-clean} and \textit{test-clean} splits of LibriAugmented~\cite{cumlin2025impairments} - an augmented version of LibriSpeech~\cite{panayotov2015librispeech} with impairments applied from the Audiomentations library\footnote{\url{https://github.com/iver56/audiomentations/}}.
Since this dataset is not publicly available, we created our own version\footnote{%
    Reproducible from \ifinterspeechfinal
        \url{https://github.com/fgnt/frame-level-mos}
    \else
        \emph{hidden during review}.
    \fi
} with two modifications.
We use background noises from CHiME-3~\cite{barker2015chime3}, and we replace the \texttt{GainTransition} impairment with the \texttt{Gain} impairment, which applies a constant gain to the full utterance and is better suited for our partial mix-up augmentation.
Each utterance is impaired exactly twice: Once with a single impairment - choosing from $9$ unique impairments - and once by applying two impairments - choosing from $6$ impairment combinations - totalling $K=15$ unique impairment classes.

\subsubsection{Model configurations}

For the encoder, we use a pretrained wav2vec2-large~\cite{hsu2021interspeech} that yields $1024$-dimensional embeddings at a frame rate of $\SI{50}{Hz}$ and that is fine-tuned alongside the decoder.
The MOS decoder is a 3-layer CNN with hidden layer sizes $256$, $256$, and $D_Z=64$, kernel sizes\footnote{The total receptive field is \SI{400}{ms}, which matches STOI's frame duration~\cite{taal2010short}.} $11$, $7$, and $5$, LeakyReLU activation, and batch normalization, followed by a linear projection to a scalar.
For $\decsclA$, we use the same configuration with a projection to $D_P=128$-dimensional embeddings, where the supervised contrastive loss is applied.
For $\mathcal{L}_{\textrm{\acrshort{lssqa}}}$, we use the hyperparameters from~\cite{kuhlmann2026speech}.
Following~\cite{khosla2020supervised}, we use $\tau=0.1$ in the computation of the supervised contrastive loss, and we set $\lambda=10$, which equals half the size of the receptive field.
\Cref{tab:model-configurations} shows an overview of the trained models.
All models are trained for $\SI{100}{epochs}$ with a batch size of $64$ and an initial learning rate of $\SI{1e-4}{}$ which is linearly decayed to $\SI{1e-6}{}$.

\begin{table}[h]
    \centering
    \setlength{\tabcolsep}{9.5pt}
    \caption{%
        Training setup.
        When training with $\mathcal{L}_{\textrm{total}}$, we consider excluding clean frames from the set of positive classes (CON2).
    }
    \vspace{-0.5em}
    \label{tab:model-configurations}
    \begin{tabular}{lcrc}
    \toprule[1.5pt]
         Model ID & Loss & $p_\textrm{mixup}$ & Include clean \\
     \midrule[1pt]
         \multicolumn{4}{l}{\footnotesize\textbf{MOS embeddings}} \\
         Baseline~\cite{kuhlmann2026speech} & $\mathcal{L}_{\textrm{\acrshort{lssqa}}}$~\labelcref{eq:loss-lssqa} & $0.0$ & n/a \\
         SUP1 & $\mathcal{L}_{\textrm{\acrshort{lssqa}}}^{\textrm{sup}}$~\labelcref{eq:loss-lssqa-sup} & $0.5$ & n/a \\
         SUP2 & $\mathcal{L}_{\textrm{\acrshort{lssqa}}}^{\textrm{sup}}$~\labelcref{eq:loss-lssqa-sup} & $1.0$ & n/a \\
     \midrule
         \multicolumn{4}{l}{\footnotesize\textbf{Contrastive embeddings}} \\
         CON1 & $\mathcal{L}_{\textrm{total}}$~\labelcref{eq:loss-supcon} & 1.0 & \yes \\
         CON2 & $\mathcal{L}_{\textrm{total}}$~\labelcref{eq:loss-supcon} & 1.0 & \no \\
     \bottomrule[1.5pt]
    \end{tabular}
    \vspace{-1em}
\end{table}

\subsubsection{Metrics}

\noindent\textbf{Local degradation detection.}
For embedding-based evaluation, we treat the similarity between the enrollment and each embedding as a single detection outcome, and report the frame-wise \gls{eer} (lower is better), and the \gls{mindcf}~\cite{przybocki2004nist}, which is a weighted sum of the miss and false-alarm rates (lower is better).
For \acrshort{mindcf} calculation, we set $p_{\textrm{target}}=0.01$, which is the assumed a priori probability of a degradation in our case.
As per-frame detection does not assess whether the detections form connected regions that have a significant intersection with the ground-truth annotation, we report the \acrlong{auc} of an intersection-based detection~(\acrshort{iauc})~\cite{bilen2020framework} for both \acrshort{mos}- and embedding-based detection (higher is better).
Intersection-based detection needs two hyperparameters: The detection tolerance criterion $\rho_{\textrm{DTC}}$
and the ground truth intersection criterion $\rho_{\textrm{GTC}}$.
We use $\rho_{\textrm{DTC}}=\rho_{\textrm{GTC}}=0.7$, which is a common choice for sound event detection~\cite{cornell2024dcase}.

\noindent\textbf{Embedding analysis.}
We perform a \emph{degradation verification} to check whether embeddings of the same degradation type are grouped together and distinguishable from other degradation types in the latent space.
We compute the pairwise distances between the embeddings and report the \acrshort{eer} of a detection using the distances to decide on the same class affiliation.
In addition to the verification task, we perform retrieval by extracting a prototype for each degradation class from the validation split and computing distances between all test embeddings and the prototypes.
For each test embedding, we take the prototype with the minimum distance and report the accuracy as the ratio of correctly selected prototypes.

\noindent\textbf{Joint detection and assignment.}
Here, we perform an agglomerative clustering (cosine distance, average linkage), assuming the number of degradation classes is known.
Each unique combination of degradations is handled as an individual class, and for \texttt{NISQA\_TEST\_SIM}-partial-mixup, all regions with more than two simultaneously active degradations are omitted from clustering to keep the number of clusters tractable.
Then, the \gls{ari} in the range \num{-0.5} to \num{1} (higher is better) is evaluated to assess the cluster purity compensated against a random assignment. 

\subsection{Local degradation detection}

\begin{table*}[t]
    \centering
    \caption{%
        Embedding- vs. MOS-based detection of degraded frames.
        \IAUC{0.7}{0.7} denotes the AUC of an intersection-based detection with $\rho_{\textrm{DTC}}=\rho_{\textrm{GTC}}=0.7$.
        Best results for each metric and embedding type are shown in \textbf{bold}.
        Overall best results per metric are additionally \underline{\textbf{underlined}}.
        Testset: \texttt{NISQA\_TEST\_SIM}-partial-mixup.
        Enrollment set: \texttt{NISQA\_VAL\_SIM}/ref.
    }
    \vspace{-0.5em}
    \label{tab:detection-val-in-domain}
    \setlength{\tabcolsep}{14pt}
    \setlength{\extrarowheight}{0.5pt}
    \begin{tabular}{%
        lr%
        r@{\extracolsep{7pt}}r@{\extracolsep{7pt}}r%
        r@{\extracolsep{7pt}}r@{\extracolsep{7pt}}r%
        r@{\extracolsep{7pt}}r@{\extracolsep{7pt}}r%
        c
    }
    \toprule[1.5pt]
        & Model ID &  \multicolumn{9}{c}{{Embedding-based}} & {{MOS-based}} \\
        \cmidrule(lr){3-11} \cmidrule(lr){12-12}
        & & \multicolumn{3}{c}{{Frame-EER [\%]}} & \multicolumn{3}{c}{{Frame-minDCF}} & \multicolumn{3}{c}{{\IAUC{0.7}{0.7}}} & {{\IAUC{0.7}{0.7}}} \\
    \midrule[1pt]
        \multirow{4}*{\begin{turn}{90}without $\mathcal{L}^{\textrm{scl}}$\end{turn}} & &
        ${\feature}$&{$\contrastembedding^{\textrm{MOS}}$}& & 
        ${\feature}$&{$\contrastembedding^{\textrm{MOS}}$}& &
        ${\feature}$&{$\contrastembedding^{\textrm{MOS}}$}& & \\
        & Baseline~\cite{kuhlmann2026speech} & \formatEERD{$19.21$}{$17.06$}{} & \formatAUCD{$1.00$}{$1.00$}{} & \formatAUCD{$0.04$}{$0.02$}{} & \formatAUC{$0.01$} \\
        & SUP1 & \formatEERD{$15.36$}{$13.34$}{} & \formatAUCD{$0.99$}{$\textbf{0.87}$}{} & \formatAUCD{$0.47$}{$0.48$}{} & \formatAUC{$0.25$} \\
        & SUP2 & \formatEERD{$12.71$}{$\textbf{11.01}$}{} & \formatAUCD{$1.00$}{$1.00$}{} & \formatAUCD{${0.57}$}{$\textbf{0.58}$}{} & \formatAUC{$0.52$} \\
    \midrule
        \multirow{3}{*}{\begin{turn}{90}with $\mathcal{L}^{\textrm{scl}}$\end{turn}} & & 
        {$\feature$}&{$\contrastembedding^{\textrm{scl}}$}&{$\tilde{\contrastembedding}$} & 
        {$\feature$}&{$\contrastembedding^{\textrm{scl}}$}&{$\tilde{\contrastembedding}$} &  
        {$\feature$}&{$\contrastembedding^{\textrm{scl}}$}&{$\tilde{\contrastembedding}$} & \\
        & CON1 & \formatEERD{$\textbf{5.22}$}{$\underline{\textbf{3.87}}$}{$\textbf{3.93}$} & \formatAUCD{$\textbf{0.97}$}{$\underline{\textbf{0.60}}$}{$\underline{\textbf{0.60}}$} & \formatAUCD{$\textbf{0.86}$}{$\underline{\textbf{0.91}}$}{$\underline{\textbf{0.91}}$} & \formatAUC{$0.65$} \\
        & CON2 & \formatEERD{$9.35$}{$5.39$}{$33.8$} & \formatAUCD{$0.99$}{$0.84$}{$1.00$} & \formatAUCD{$0.50$}{$0.81$}{$0.10$} & \formatAUC{$\underline{\textbf{0.67}}$} \\
    \bottomrule[1.5pt]
    \end{tabular}
    \vspace{0.5em}
\end{table*}

We perform \acrshort{mos}- and embedding-based detection as described in~\Cref{sec:localization}.
Note that our metrics are threshold-independent, so we do not tune any thresholds on the validation split.
For embedding-based detection, we use the validation splits as the enrollment set.
To compute the enrollment embedding, we compute an utterance embedding as the mean of the frame embeddings and average all utterance embeddings.

\Cref{tab:detection-val-in-domain} shows the results for the in-domain test data.
Adding frame-level pseudo labels to the training significantly improves the performance of \acrshort{mos}-based detection (Baseline \acrshort{iauc}: $0.01$, SUP2 \acrshort{iauc}: $0.52$).
Higher mix-up ratios lead to better results (SUP1 \acrshort{iauc}: $0.25$, SUP2 \acrshort{iauc}: $0.52$). 
The embedding-based detection for the \acrshort{mos}-only models also profits from the supervision (Baseline $\vect{x}$-\acrshort{iauc}: $0.04$ versus SUP2 $\vect{x}$-\acrshort{iauc}: $0.57$).
Note that in \acrshort{sed}, an \acrshort{iauc} of around $0.6$ is considered state-of-the-art~\cite{ebbers2023post}.
However, detection costs are high with the lowest \acrshort{mindcf} of $0.87$ achieved by SUP1.

\begin{figure}[h]
    \centering
    \includegraphics[width=\columnwidth]{%
        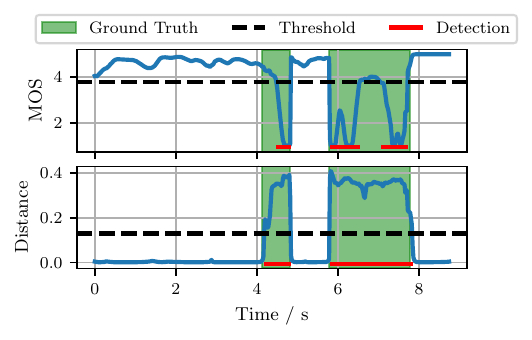
    }
    \vspace{-2.25em}
    \caption{%
        CON1 \acrshort{mos}-based (upper) and embedding-based (lower) detection results for a single utterance from \texttt{NISQA\_TEST\_SIM}-partial-mixup.
        The threshold is tuned to maximize the intersection-based F1-score on the full test set.
    }
    \label{fig:detection-con1}
\end{figure}

Training with the supervised contrastive loss leads to a significant drop in \acrshort{eer} (SUP2: $11.1\%$, CON1: $3.87\%$) and a significant increase in \acrshort{iauc} for embedding-based detection (SUP2: $0.58$, CON1: $0.91$).
Including clean frames as a distinct positive class yields a significant performance improvement compared to excluding them (e.g., CON1 \acrshort{mindcf}: $0.60$, CON2 \acrshort{mindcf}: $0.84$).
Regarding the embedding choice, the decoder bottleneck embeddings $\contrastembedding^{\textrm{scl}}$ perform most robustly.
The supervised contrastive loss slightly improves \acrshort{mos}-based detection performance (CON2 best \acrshort{mos}-\acrshort{iauc} of $0.67$); however, a large gap to embedding-based detection remains (CON1 best embedding-\acrshort{iauc} of $0.91$); hence, embedding-based detection should be preferred.
\Cref{fig:detection-con1} shows a comparison between \acrshort{mos}- and embedding-based detection on a single example.
The better performance of the embedding-based detection stems from a larger margin to the threshold for both clean and degraded frames, and from higher sensitivity to degradation frame onsets (here for the first degraded segment).

\begin{table*}[t]
    \centering
    \caption{%
        Embedding- vs. MOS-based detection on unseen degradations.
        Best results for each metric and embedding type are shown in \textbf{bold}.
        Overall best results per metric are additionally \underline{\textbf{underlined}}.
        Testset: LibriAugmented/test-clean-partial-mixup.
        Enrollment set: LibriTTS/dev-clean.
    }
    \vspace{-0.5em}
    \label{tab:detection-val-out-of-domain}
    \setlength{\tabcolsep}{13.2pt}
    \setlength{\extrarowheight}{0.5pt}
    \begin{tabular}{%
        lr%
        r@{\extracolsep{7pt}}r@{\extracolsep{7pt}}r%
        r@{\extracolsep{7pt}}r@{\extracolsep{7pt}}r%
        r@{\extracolsep{7pt}}r@{\extracolsep{7pt}}r%
        c
    }
    \toprule[1.5pt]
        & Model ID &  \multicolumn{9}{c}{{Embedding-based}} & {{MOS-based}} \\
        \cmidrule(lr){3-11} \cmidrule(lr){12-12}
        & & \multicolumn{3}{c}{{Frame-EER [\%]}} & \multicolumn{3}{c}{{Frame-minDCF}} & \multicolumn{3}{c}{{\IAUC{0.7}{0.7}}} & {{\IAUC{0.7}{0.7}}} \\
    \midrule[1pt]
        \multirow{4}*{\begin{turn}{90}without $\mathcal{L}^{\textrm{scl}}$\end{turn}} & &
        ${\feature}$&{$\contrastembedding^{\textrm{MOS}}$}& & 
        ${\feature}$&{$\contrastembedding^{\textrm{MOS}}$}& &
        ${\feature}$&{$\contrastembedding^{\textrm{MOS}}$}& & \\
        & Baseline~\cite{kuhlmann2026speech} & \formatEERD{15.29}{16.61}{} & \formatAUCD{1.00}{1.00}{} & \formatAUCD{0.30}{0.07}{} & \formatAUC{0.20} \\
        & SUP1 & \formatEERD{13.34}{{12.16}}{} & \formatAUCD{\textbf{0.93}}{\textbf{0.93}}{} & \formatAUCD{0.63}{0.43}{} & \formatAUC{0.59} \\
        & SUP2 & \formatEERD{11.07}{\textbf{10.79}}{}  & \formatAUCD{0.95}{1.00}{} & \formatAUCD{0.79}{\textbf{0.72}}{} & \formatAUC{0.76} \\
    \midrule
        \multirow{3}{*}{\begin{turn}{90}with $\mathcal{L}^{\textrm{scl}}$\end{turn}} & & 
        {$\feature$}&{$\contrastembedding^{\textrm{scl}}$}&{$\tilde{\contrastembedding}$} & 
        {$\feature$}&{$\contrastembedding^{\textrm{scl}}$}&{$\tilde{\contrastembedding}$} &  
        {$\feature$}&{$\contrastembedding^{\textrm{scl}}$}&{$\tilde{\contrastembedding}$} & \\
        & CON1 & \formatEERD{\textbf{7.32}}{{\textbf{4.88}}}{\underline{\textbf{4.64}}} & \formatAUCD{1.00}{{\textbf{0.62}}}{\underline{\textbf{0.61}}} & \formatAUCD{\textbf{0.86}}{\underline{\textbf{0.92}}}{\underline{\textbf{0.92}}} & \formatAUC{0.77} \\
        & CON2 & \formatEERD{9.51}{7.60}{26.7} & \formatAUCD{1.00}{0.95}{1.00} & \formatAUCD{0.78}{0.80}{0.05} & \formatAUC{\underline{\textbf{0.78}}} \\
    \bottomrule[1.5pt]
    \end{tabular}
    \vspace{-0.5em}
\end{table*}

\Cref{tab:detection-val-out-of-domain} shows the result on out-of-domain degradations.
We observe the same trends and improvements as in the in-domain case, with slightly better \acrshort{mos}-based detection (CON2-\acrshort{iauc}: 0.78 versus 0.67), but slightly worse \acrshort{eer} (CON1 $\contrastembedding^\textrm{scl}$-EER: 4.88\% versus 3.87\%).

\subsection{Embedding analysis}
\label{sec:embedding-analysis}

To analyze whether frame-level embeddings of the same degradation type group together and are distinguishable from embeddings of other degradation types, we use the ground-truth annotations to detect degraded embeddings.
We evaluate the embedding purity when using the actual detections from the embedding-based detection in~\Cref{sec:joint}.
Given the usually strong correlation of neighboring frames,
we reduce the number of frame embeddings to analyze by taking the mean of consecutively detected embeddings.

\begin{table}[h]
    \centering
    \caption{%
        Performance of degradation-type verification and retrieval for all $K=36$ degradation combinations, and using oracle information to detect the degradations.
        Overall best results for each metric are shown in \textbf{bold}.
        Testset: \texttt{NISQA\_TEST\_SIM}-partial-mixup.
    }
    \vspace{-0.5em}
    \label{tab:embedding-analysis-nisqa-val-sim}
    \setlength{\tabcolsep}{12.5pt}
    \begin{tabular}{%
        r
        r@{\extracolsep{7pt}}r
        r@{\extracolsep{7pt}}r
    }
    \toprule[1.5pt]
        Model ID & \multicolumn{2}{c}{EER [\%]} & \multicolumn{2}{c}{Accuracy [\%]} \\
    \midrule[1pt]
        & 
        $\feature$&$\contrastembedding^{\textrm{MOS}}$ & 
        $\feature$&$\contrastembedding^{\textrm{MOS}}$ \\
        Baseline~\cite{kuhlmann2026speech} & \formatEERB{42.11}{43.86} & \formatEERB{2.12}{0.62} \\
        SUP1 & \formatEERB{37.81}{39.11} & \formatEERB{0.61}{0.62} \\
        SUP2 & \formatEERB{36.29}{44.64} & \formatEERB{7.14}{2.12} \\
    \midrule
        &
        $\contrastembedding^{\textrm{scl}}$&$\tilde{\contrastembedding}$ & 
        $\contrastembedding^{\textrm{scl}}$&$\tilde{\contrastembedding}$ \\
        CON1 & \formatEERB{15.17}{14.15} & \formatEERB{\textbf{30.30}}{27.41} \\
        CON2 & \formatEERB{\textbf{13.56}}{14.25} & \formatEERB{26.79}{25.76} \\
    \bottomrule[1.5pt]
    \end{tabular}
\end{table}

\Cref{tab:embedding-analysis-nisqa-val-sim} shows the results on \texttt{NISQA\_TEST\_SIM}-partial-mixup when including all $36$ degradation combinations.
Adding frame-level pseudo-targets to the training does not drastically improve discrimination (Baseline \acrshort{eer}: $42.1\%$, SUP2 \acrshort{eer}: $36.3\%$).
Training with $\mathcal{L}_{\textrm{total}}$ significantly improves \acrshort{eer} (CON2: $13.5\%$) and accuracy (SUP2: $7.14\%$, CON1: $30.3\%$).
Excluding clean frames from the loss gives only marginal improvement in \acrshort{eer} (CON1: $15.2\%$, CON2: $13.5\%$).

\begin{figure}[h]
    \centering
    \includegraphics[width=\linewidth]{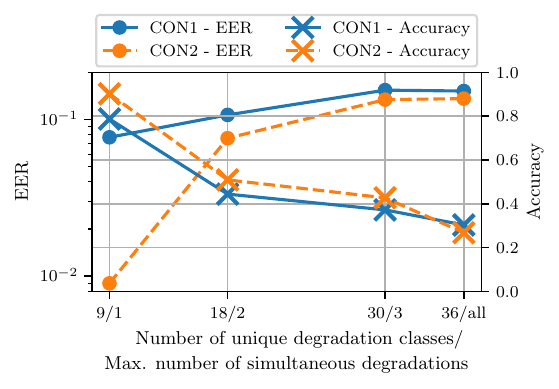}
    \vspace{-2.25em}
    \caption{%
        \acrshort{eer} and accuracy when constraining the number of concurrent degradations in \texttt{NISQA\_TEST\_SIM}-partial-mixup.
     }
    \label{fig:eer-over-number-classes}
\end{figure}

\Cref{fig:eer-over-number-classes} shows how \acrshort{eer} and accuracy for CON1 and CON2 behave when constraining the number of simultaneous degradations and thus the number of unique degradation classes.
Single degradations ($K=9$) can be distinguished accurately (CON1 \acrshort{eer}: $7.71\%$, CON2 \acrshort{eer}: $0.93\%$).
Here, it becomes evident that excluding clean frames from the set of positive classes during training (CON2) is clearly beneficial for degradation identification.
When increasing the maximum number of simultaneous degradations, accuracy starts to drop and \acrshort{eer} gets higher, as more classes (two degradations: $K=18$; three degradations: $K=30$) share identical distortions, leading to more confusion between classes with shared degradations.

\begin{table}[h]
    \centering
    \caption{%
        Performance of degradation-type verification and retrieval for unseen degradations using oracle information to detect the degradations.
        We also report results when restricting to single degradations.
        Overall best results for each metric and number of simultaneous degradations are shown in \textbf{bold}.
        Testset: LibriAugmented/test-clean-partial-mixup.
    }
    \vspace{-0.5em}
    \label{tab:embedding-analysis-libritts-dev-clean}
    \setlength{\tabcolsep}{3pt}
    \begin{tabular}{
        r
        l@{\extracolsep{2pt}}l l@{\extracolsep{2pt}}l
        l@{\extracolsep{2pt}}l l@{\extracolsep{2pt}}l
    }
    \toprule[1.5pt]
        & \multicolumn{4}{c}{Single} & \multicolumn{4}{c}{All} \\
        \cmidrule(lr){2-5} \cmidrule(lr){6-9}
        Model ID & \multicolumn{2}{c}{EER [\%]} & \multicolumn{2}{c}{Acc. [\%]} & \multicolumn{2}{c}{EER [\%]} & \multicolumn{2}{c}{Acc. [\%]} \\
    \midrule[1pt]
        & $\feature$&$\contrastembedding^{\textrm{MOS}}$ 
        & $\feature$&$\contrastembedding^{\textrm{MOS}}$ 
        & $\feature$&$\contrastembedding^{\textrm{MOS}}$ 
        & $\feature$&$\contrastembedding^{\textrm{MOS}}$ \\
        Baseline~\cite{kuhlmann2026speech} & \formatEERB{38.0}{38.7} & \formatEERB{39.7}{36.7} & \formatEERB{36.2}{38.6} & \formatEERB{32.5}{27.6} \\
        SUP1 & \formatEERB{33.4}{36.5} & \formatEERB{46.2}{43.7} & \formatEERB{35.2}{37.9} & \formatEERB{34.8}{27.4} \\
        SUP2 & \formatEERB{33.6}{39.7} & \formatEERB{46.8}{40.3} & \formatEERB{33.9}{38.5} & \formatEERB{35.2}{28.3} \\
    \midrule
        & $\contrastembedding^{\textrm{scl}}$&$\tilde{\contrastembedding}$ & 
        $\contrastembedding^{\textrm{scl}}$&$\tilde{\contrastembedding}$ & 
        $\contrastembedding^{\textrm{scl}}$&$\tilde{\contrastembedding}$ & 
        $\contrastembedding^{\textrm{scl}}$&$\tilde{\contrastembedding}$ \\
        CON1 & \formatEERB{{19.4}}{\textbf{19.2}} &  \formatEERB{77.5}{\textbf{78.9}} & \formatEERB{{19.7}}{\textbf{19.5}} &  \formatEERB{65.0}{\textbf{65.6}} \\
        CON2 & \formatEERB{20.4}{20.6} & \formatEERB{{77.2}}{76.1} & \formatEERB{19.8}{19.9} & \formatEERB{65.1}{64.5} \\
    \bottomrule[1.5pt]
    \end{tabular}
    \vspace{0.5em}
\end{table}

Regarding the results on out-of-domain degradations (\Cref{tab:embedding-analysis-libritts-dev-clean}), again, significant improvements can be gained when training with the supervised contrastive loss (SUP2 accuracy: $46.8\%$/$35.2\%$, CON1 accuracy: $78.9\%$/$65.6\%$).
While accuracy behaves similarly to the in-domain case, the \acrshort{eer} is much higher at around $19\%$, indicating that unseen degradations are harder to separate from one another.

\begin{table}
    \centering
        \caption{%
            Clustering results in terms of \acrfull{ari} 
            and ARI$^{\text{dist}}$ (without the clean speech class).
            ACC denotes the accuracy for the clean speech class for clustering.
        }
        \vspace{-0.5em}
        \sisetup{round-mode=places,round-precision=2}
    \label{tab:clustering} 
    \setlength{\tabcolsep}{2.5pt}
    \begin{tabular}{l c S S S S S S }
    \toprule[1.5pt]
        Model ID & Detec. & \multicolumn{3}{c}{{NISQA}} & \multicolumn{3}{c}{LibriAugmented} \\
        \cmidrule(lr){3-5}\cmidrule(lr){6-8}
        & &{ARI$^{\text{dist}}$} & {ARI} & {ACC} & {ARI$^{\text{dist}}$} & {ARI} &  {ACC} \\
        \midrule[1pt]
         CON1 & orc. & 
         0.5954446341725323 & 
         {--} & 
         {--} & 
         \B 0.2844498133676942 &
         {--} &
         {--} \\ 
         CON2 & orc. & 
         \B 0.7288347759916294 & 
         {--} & 
         {--} & 
         \B 0.27354226325491293 &
         {--} & 
         {--} \\
         \midrule 
         CON1 & CON1 &
         0.4336310726216035 &
         \B .676260739741207 &
         \B 0.8953846153846153 &
         0.23584105390704274 &
         0.4750658518236747 &
         \B 0.9275082690187431 \\
         CON2 & CON2 &
         0.4751174350836737 &
         0.2686508091402248 &
         0.14492753623188406 &
         \B 0.2934757983730547 &
         0.30895547608290264 &
         0.42597500946611133 \\
         \midrule
         CON2 & CON1 &
         0.5629411464348143 &
         0.27919140155689853 &
         0.6476683937823834 &
         \B 0.278292535324361 &
         \B 0.7821715108115478 &
         \B 0.9375689084895259
\\ 
     \bottomrule[1.5pt]
    \end{tabular}
\vspace{0.5em}
\end{table}

\subsection{Joint degradation detection and assignment}
\label{sec:joint}
So far, the tasks of degradation detection and degradation-type grouping have been treated separately.
However, for a downstream application, both are required. 
Thus, the latent embedding structures of the two contrastive models, CON1 and CON2, are compared.
\Cref{tab:clustering} shows the comparison with and without oracle detection. 
Here, additionally, the \gls{ari} while ignoring the clean speech class is evaluated ({ARI$^{\text{dist}}$), and the accuracy of the clean speech cluster (ACC) is shown.
On the in-domain data, for an oracle segmentation, the model trained without the clean speech positive class (CON2) proves significantly better, as also seen in \Cref{sec:embedding-analysis}.
However, when using a non-oracle detection, this model significantly degrades. 
When comparing the \gls{ari} with and without scoring the clean speech cluster, this difference is due to the model's inability to consistently output embeddings for clean speech (\gls{ari} of $0.27$ versus $\textrm{\gls{ari}}^{{\textrm{dist}}}$ of $0.48$). 
In contrast, for model CON1, the clean-speech embeddings are accurately grouped into a single cluster.
Using the model CON1 for detection while extracting embeddings with model CON2 mitigates this behavior, but the mismatch between clustering performance persists, so that only two third of the clean-speech embeddings are correctly clustered.

For the out-of-domain data, overall clustering accuracy drops.
The lower \gls{ari} indicates that the model cannot fully separate different degradation types, consistent with the high \glspl{eer} in~\Cref{tab:embedding-analysis-libritts-dev-clean}. 
However, model CON1 still allows accurate discrimination between false positives (clean speech) and degraded embeddings (clean cluster accuracy: $93\%$).
We hypothesize that the drop in clustering performance is due to an inconsistent embedding behavior for degradations unseen during training, but leave a more in-depth analysis for future work.
For stability, the inclusion of clean-speech embeddings during training is mandatory to prevent the model from degrading under non-ideal detection.
Then, the model yields well-defined clusters for both seen degradations and clean speech. For unseen degradations, this separation is less pronounced, 
but the embeddings are clearly separate from clean-speech embeddings.

\section{Conclusion}
We have shown that the detection of local degradation in speech signals can be significantly improved using (i) frame-level pseudo-targets via a partial mix-up data augmentation, (ii) adding a supervised contrastive loss exploiting knowledge about the degradation types, and (iii) switching from \acrshort{mos}-based to embedding-based detection, resulting in near-perfect detection performance on in- and out-of-domain data.
The contrastive loss also improved discrimination between degradation types and retrieval of the same types.
Further analysis is needed to improve degradation-type discrimination on in- and out-of-domain data and to assess robustness under varying levels of degradation~\cite {deng2025investigating}.
To this end, an analysis of the similarity among degradation types would be highly beneficial for creating more representative training and test data.
Finally, this system can be evolved along the same lines as frame-level audio-language models~\cite{wu2025flam,primus2025tacos} and trained on textual descriptions to enable automatic assignment of frame embeddings or clusters to degradation types.
\section{Acknowledgements}

\ifinterspeechfinal
     Computational resources were provided by the Paderborn Center for Parallel Computing.
\else
     Acknowledgements hidden during review.
\fi

\bibliographystyle{IEEEtran}
\bibliography{references}

@string{icassp = "Proceedings of International Conference on Acoustics, Speech and Signal Processing (ICASSP)"}

@string{interspeech = "Proceedings of ISCA Interspeech"}

@string{dcase = "Workshop on Detection and Classification of Acoustic Scenes and Events (DCASE)"}

@string{icml = "Proceedings of the International Conference on Machine Learning (ICML)"}

@string{waspaa = "Proceedings of Workshop on Applications of Signal Processing to Audio and Acoustics (WASPAA)"}

@string{asru = "Proceedings of Automatic Speech Recognition and Understanding Workshop (ASRU)"}

@string{odyssey = "Speaker and Language Recognition Workshop (Odyssey)"}

@string{nips = "Advances in Neural Information Processing Systems (NeurIPS)"}

@inproceedings{rix2001perceptual,
  title={Perceptual evaluation of speech quality (PESQ)-a new method for speech quality assessment of telephone networks and codecs},
  author={Rix, Antony W and Beerends, John G and Hollier, Michael P and Hekstra, Andries P},
  booktitle=icassp,
  pages={749--752},
  year={2001},
}

@inproceedings{przybocki2004nist,
  title     = {{NIST speaker recognition evaluation chronicles}},
  author    = {Mark Przybocki and Alvin F. Martin},
  year      = {2004},
  booktitle = odyssey,
  pages     = {15--22},
}

@inproceedings{taal2010short,
  title={A short-time objective intelligibility measure for time-frequency weighted noisy speech},
  author={Taal, Cees H and Hendriks, Richard C and Heusdens, Richard and Jensen, Jesper},
  booktitle=icassp,
  pages={4214--4217},
  year={2010},
}

@article{hines2012speech,
  title={Speech intelligibility prediction using a neurogram similarity index measure},
  author={Hines, Andrew and Harte, Naomi},
  journal={Speech Communication},
  volume={54},
  number={2},
  pages={306--320},
  year={2012},
  publisher={Elsevier}
}

@INPROCEEDINGS{panayotov2015librispeech,
  author={Panayotov, Vassil and Chen, Guoguo and Povey, Daniel and Khudanpur, Sanjeev},
  booktitle=icassp, 
  title={{Librispeech: An ASR corpus based on public domain audio books}}, 
  year={2015},
  pages={5206-5210},
  doi={10.1109/ICASSP.2015.7178964}
}

@inproceedings{barker2015chime3,
  title={The third {'CHiME’ speech} separation and recognition challenge: Dataset, task and baselines},
  author={Barker, Jon and Marxer, Ricard and Vincent, Emmanuel and Watanabe, Shinji},
  booktitle=asru,
  pages={504--511},
  year={2015},
}

@inproceedings{snyder2018x,
  title={{X-Vectors: Robust DNN Embeddings for Speaker Recognition}},
  author={Snyder, David and Garcia-Romero, Daniel and Sell, Gregory and Povey, Daniel and Khudanpur, Sanjeev},
  booktitle=icassp,
  pages={5329--5333},
  year={2018},
}

@inproceedings{fu2018qualitynet,
  title     = {{Quality-Net: An End-to-End Non-intrusive Speech Quality Assessment Model Based on BLSTM}},
  author    = {Szu-wei Fu and Yu Tsao and Hsin-Te Hwang and Hsin-Min Wang},
  year      = {2018},
  booktitle = interspeech}

@ARTICLE{fu2019learning,
  author={Fu, Szu-Wei and Liao, Chien-Feng and Tsao, Yu},
  journal={icassp}, 
  year={2020},
  pages={26-30},
}

@InProceedings{fu2019metric,
  title = 	 {{M}etric{GAN}: Generative Adversarial Networks based Black-box Metric Scores Optimization for Speech Enhancement},
  author =       {Fu, Szu-Wei and Liao, Chien-Feng and Tsao, Yu and Lin, Shou-De},
  booktitle = 	 {Proceedings of the International Conference on Machine Learning},
  pages = 	 {2031--2041},
  year = 	 {2019},
}

@inproceedings{turpault2019sed,
  title={{Sound Event Detection in Domestic Environments with Weakly Labeled Data and Soundscape Synthesis}},
  author={Nicolas Turpault and Romain Serizel and Ankit Shah and Justin Salamon},
  booktitle=dcase,
  year={2019},
}

@inproceedings{chinen2020visqol,
  title={ViSQOL v3: An open source production ready objective speech and audio metric},
  author={Chinen, Michael and Lim, Felicia SC and Skoglund, Jan and Gureev, Nikita and O'Gorman, Feargus and Hines, Andrew},
  booktitle={International Conference on Quality of Multimedia Experience (QoMEX)},
  pages={1--6},
  year={2020},
}

@article{khosla2020supervised,
  title={Supervised contrastive learning},
  author={Khosla, Prannay and Teterwak, Piotr and Wang, Chen and Sarna, Aaron and Tian, Yonglong and Isola, Phillip and Maschinot, Aaron and Liu, Ce and Krishnan, Dilip},
  journal=nips,
  pages={18661--18673},
  year={2020}
}

@article{tan2020rvad,
  title={{rVAD: An unsupervised segment-based robust voice activity detection method}},
  author={Tan, Zheng-Hua and Dehak, Najim and others},
  journal={Computer Speech \& Language},
  pages={1--21},
  year={2020},
  publisher={Elsevier}
}

@article{bilen2020framework,
  title={{A Framework for the Robust Evaluation of Sound Event Detection}},
  author={Çağdaş Bilen and Giacomo Ferroni and Francesco Tuveri and Juan Azcarreta and Sacha Krstulovic},
  journal=icassp,
  year={2020},
  pages={61-65},
}

@inproceedings{chung2020defence,
  title={In Defence of Metric Learning for Speaker Recognition},
  author={Chung, Joon Son and Huh, Jaesung and Mun, Seongkyu and Lee, Minjae and Heo, Hee-Soo and Choe, Soyeon and Ham, Chiheon and Jung, Sunghwan and Lee, Bong-Jin and Han, Icksang},
  booktitle=interspeech,
  pages={2977--2981},
  year={2020}
}

@article{cooper2021generalization,
  title={{Generalization Ability of MOS Prediction Networks}},
  author={Erica Cooper and Wen-Chin Huang and Tomoki Toda and Junichi Yamagishi},
  journal=icassp,
  year={2022},
  pages={8442-8446},
}

@article{leng2021mbnet,
  title={{MBNET: MOS Prediction for Synthesized Speech with Mean-Bias Network}},
  author={Yichong Leng and Xu Tan and Sheng Zhao and Frank K. Soong and Xiang-Yang Li and Tao Qin},
  journal=icassp,
  year={2021},
  pages={391-395},
}

@inproceedings{reddy2021dnsmos,
  title={DNSMOS: A non-intrusive perceptual objective speech quality metric to evaluate noise suppressors},
  author={Reddy, Chandan KA and Gopal, Vishak and Cutler, Ross},
  booktitle=icassp,
  pages={6493--6497},
  year={2021},
}

@inproceedings{mittag2021nisqa,
  title     = {{NISQA: A Deep CNN-Self-Attention Model for Multidimensional Speech Quality Prediction with Crowdsourced Datasets}},
  author    = {Gabriel Mittag and Babak Naderi and Assmaa Chehadi and Sebastian Möller},
  year      = {2021},
  booktitle = interspeech,
  pages     = {2127--2131},
  doi       = {10.21437/Interspeech.2021-299},
  issn      = {2958-1796},
}

@inproceedings{cooper2021how,
  title     = {How do Voices from Past Speech Synthesis Challenges Compare Today?},
  author    = {Erica Cooper and Junichi Yamagishi},
  year      = {2021},
  booktitle = {11th ISCA Speech Synthesis Workshop},
  pages     = {183--188},
  doi       = {10.21437/SSW.2021-32},
}

@inproceedings{hsu2021interspeech,
  title     = {{Robust wav2vec 2.0: Analyzing Domain Shift in Self-Supervised Pre-Training}},
  author    = {Wei-Ning Hsu and Anuroop Sriram and Alexei Baevski and Tatiana Likhomanenko and Qiantong Xu and Vineel Pratap and Jacob Kahn and Ann Lee and Ronan Collobert and Gabriel Synnaeve and Michael Auli},
  year      = {2021},
  booktitle = interspeech,
  pages     = {721--725},
  doi       = {10.21437/Interspeech.2021-236},
  issn      = {2958-1796},
}

@article{hao2022dcase,
  title={{DCASE 2022 TASK4 CHALLENGE TECHNICAL REPORT}},
  author={Hao, Junyong and Ye, Shunzhou and Lu, Cheng and Dong, Fei and Liu, Jingang},
  journal=dcase,
  year={2022}
}

@inproceedings{saeki2022utmos,
  title     = {{UTMOS: UTokyo-SaruLab System for VoiceMOS Challenge 2022}},
  author    = {{Takaaki Saeki and Detai Xin and Wataru Nakata and Tomoki Koriyama and Shinnosuke Takamichi and Hiroshi Saruwatari}},
  year      = {{2022}},
  booktitle = interspeech,
  pages     = {{4521--4525}},
  doi       = {{10.21437/Interspeech.2022-439}},
  issn      = {{2958-1796}},
}

@article{chen2023dcase,
  title={{DCASE 2023 Challenge Task4 Technical Report}},
  author={Chen, Minjun and Jin, Yongbin and Shao, Jun and Liu, Yangyang and Peng, Bo and Chen, Jie},
  journal=dcase,
  year={2023}
}

@article{tian2023stablerep,
  title={Stablerep: Synthetic images from text-to-image models make strong visual representation learners},
  author={Tian, Yonglong and Fan, Lijie and Isola, Phillip and Chang, Huiwen and Krishnan, Dilip},
  journal=nips,
  pages={48382--48402},
  year={2023}
}

@inproceedings{ebbers2023post,
  title={{Post-Processing Independent Evaluation of Sound Event Detection Systems}},
  author={Ebbers, Janek and Haeb-Umbach, Reinhold and Serizel, Romain},
  booktitle=dcase,
  year={2023}
}

@article{cornell2024dcase,
  title={{DCASE 2024 task 4: Sound event detection with heterogeneous data and missing labels}},
  author={Cornell, Samuele and Ebbers, Janek and Douwes, Constance and Mart{\'\i}n-Morat{\'o}, Irene and Harju, Manu and Mesaros, Annamaria and Serizel, Romain},
  journal={arXiv preprint arXiv:2406.08056},
  year={2024}
}

@INPROCEEDINGS{ragano2024nomad,
  author={Ragano, Alessandro and Skoglund, Jan and Hines, Andrew},
  title={{NOMAD: Unsupervised Learning of Perceptual Embeddings For Speech Enhancement and Non-Matching Reference Audio Quality Assessment}}, 
  booktitle=icassp,
  year={2024},
  pages={1011-1015},
}

@inproceedings{cumlin2025impairments,
  title={Impairments are Clustered in Latents of Deep Neural Network-based Speech Quality Models},
  author={Cumlin, Fredrik and Liang, Xinyu and Ungureanu, Victor and Reddy, Chandan KA and Sch{\"u}ldt, Christian and Chatterjee, Saikat},
  booktitle=icassp,
  pages={1--5},
  year={2025},
}

@inproceedings{deng2025investigating,
  title={Investigating the sensitivity of pre-trained audio embeddings to common effects},
  author={Deng, Victor and Wang, Changhong and Richard, Gael and McFee, Brian},
  booktitle=icassp,
  pages={1--5},
  year={2025},
}

@inproceedings{huang2025sheet,
  title     = {{SHEET: A Multi-purpose Open-source Speech Human Evaluation Estimation Toolkit}},
  author    = {Wen-Chin Huang and Erica Cooper and Tomoki Toda},
  year      = {2025},
  booktitle = interspeech,
  pages     = {2355--2359},
  doi       = {10.21437/Interspeech.2025-1977},
  issn      = {2958-1796},
}

@inproceedings{kuhlmann2025towards,
  title     = {{Towards Frame-level Quality Predictions of Synthetic Speech }},
  author    = {Michael Kuhlmann and Fritz Seebauer and Petra Wagner and Reinhold Haeb-Umbach},
  year      = {2025},
  booktitle = interspeech,
  pages     = {2300--2304},
  doi       = {10.21437/Interspeech.2025-2190},
  issn      = {2958-1796},
}

@inproceedings{wu2025flam,
    title={{FLAM}: Frame-Wise Language-Audio Modeling},
    author={Yusong Wu and Christos Tsirigotis and Ke Chen and Cheng-Zhi Anna Huang and Aaron Courville and Oriol Nieto and Prem Seetharaman and Justin Salamon},
    booktitle=icml,
    year={2025},
    url={https://openreview.net/forum?id=7fQohcFrxG}
}

@inproceedings{primus2025tacos,
  title={Tacos: Temporally-aligned audio captions for language-audio pretraining},
  author={Primus, Paul and Schmid, Florian and Widmer, Gerhard},
  booktitle=waspaa,
  pages={1--5},
  year={2025},
}

@article{kuhlmann2026speech,
  title={Speech Quality-Based Localization of Low-Quality Speech and Text-to-Speech Synthesis Artefacts},
  author={Kuhlmann, Michael and Werning, Alexander and von Neumann, Thilo and Haeb-Umbach, Reinhold},
  journal=icassp,
  year={2026},
}

\end{document}